# Thermoelectric signature of quantum criticality in the heavy-fermion superconductor CeRhIn$_5$

Zi-Yu Cao,[1, 2, *] Honghong Wang,[1, *, †] Chan-Koo Park,[1] Tae Beom Park,[1, 3] Harim Jang,[1] Soonbeom Seo,[1] Sung-Il Kim,[1] and Tuson Park[1, †]

[1] *Center for Quantum Materials and Superconductivity (CQMS) and Department of Physics, Sungkyunkwan University, Suwon 16419, Republic of Korea*

[2] *School of Physics Science & Information Technology, Liaocheng University, Liaocheng 252059, China*

[3] *Institute of Basic Science, Sungkyunkwan University, Suwon 16419, South Korea*

† Correspondence should be addressed to TP and HW (tp8701@skku.edu and hw8793@gmail.com)
* These authors have contributed equally to this work.

**The evolution of the Fermi surface across the quantum critical point (QCP), which is relevant for characterizing the quantum criticality and understanding its relation with unconventional superconductivity, is an intriguing subject in the study of strongly correlated electron systems. In this study, we report the thermopower measurements to investigate a change in Fermi surface across the QCP in pure and 4.4% Sn-doped CeRhIn$_5$. Results show that their thermopower behavior differs significantly in the vicinity of their respective pressure-induced QCP. In pure CeRhIn$_5$, a drastic collapse of the thermopower takes place at the Kondo breakdown QCP, where the Fermi surface reconstructs concurrently with the development of the magnetic order. By contrast, the thermopower exhibits a broadly symmetric behavior around the QCP in 4.4% Sn-doped CeRhIn$_5$, which is a characteristic of the spin-density-wave QCP. These observations are consistent with the theoretical expectations and suggest the effectiveness of thermopower measurement in discriminating the nature of quantum criticality in heavy-fermion systems.**

The Fermi-surface change associated with the electronic localization–delocalization transition is emerging as a unifying theme across the strongly correlated electron materials [1-3]. Among them, the localization–delocalization transition of *f*-electrons in the heavy-fermion system is realized by the destruction of the Kondo effect. When the Kondo effect is destroyed at the magnetic quantum critical point (QCP) where the magnetic boundary reaches zero temperature, i.e., the delocalization of *f*-electrons and the concomitant reconstruction of the Fermi surface from small to large coincide with the onset of magnetic order, the nature of the magnetic QCP is of the Kondo-breakdown type [4,5]. Consequently, the quantum criticality around a Kondo-breakdown QCP incorporates not only the spin fluctuations from the suppression of the magnetic-order parameter but also the emergent charge fluctuations associated with the destruction of the Kondo effect [6-9]. This is contrary to the conventional quantum criticality within the Landau framework, which is of the spin-density-wave (SDW) type and involves only the quantum critical fluctuations of the magnetic-order parameter [10].

CeRhIn$_5$ is a prototypical quantum-critical heavy-fermion system whose antiferromagnetic (AFM) transition is continuously suppressed from 3.8 K at ambient pressure to zero temperature at a critical pressure $P_c$ = 2.3 GPa, thereby accessing a pressure-induced AFM QCP [11,12]. Evidence for an abrupt reconstruction of the Fermi surface across the QCP has been inferred from the de Haas–van Alphen measurements [13], indicative of the nature of the Kondo-breakdown quantum criticality. Another signature for the Kondo-breakdown quantum criticality has come from the Hall measurements, in which the charge delocalization crossover scale, characteristic of the Kondo breakdown, terminates at the QCP [14]. Similar analysis of the Hall data for CeRh(In$_{0.956}$Sn$_{0.044}$)$_5$, labeled as Sn-doped CeRhIn$_5$ hereinafter, finds that substituting Sn for In shifts the charge delocalization crossover scale such that it intersects the AFM phase boundary at a finite temperature; thus, the Kondo breakdown is decoupled from the AFM QCP, and the corresponding AFM QCP is of the SDW type [14]. With a simple interpretation of the Hall effect, a step-like jump in carrier density is anticipated as the Fermi surface changes from small to large at $P_c$ in CeRhIn$_5$, as revealed in another representative Kondo-breakdown quantum critical material YbRh$_2$Si$_2$ across the field-induced QCP [1,15]. However, the pressure-dependent isothermal Hall coefficient peaks strongly at $P_c$ in CeRhIn$_5$ and exhibits a broad maximum in Sn-doped CeRhIn$_5$ [14]. The Hall effect in a compensated heavy-fermion metal like CeRhIn$_5$ cannot be explained merely by the changes in carrier density, implying that the low-temperature Hall coefficient cannot be used to determine the Fermi-surface volume. To further investigate the evolution of the Fermi surface as a function of pressure, especially in Sn-doped CeRhIn$_5$, experiments that can probe the Fermi surface are required.

Using the Boltzmann equation, the thermopower $S$ (also known as the Seebeck coefficient) can be expressed by the Mott formula [16], $S/T = -(\pi^2/3)(k_B^2/e)[\partial \ln\sigma(\varepsilon)/\partial\varepsilon]_{\varepsilon=\varepsilon_F}$, where $k_B$, e, $\sigma(\varepsilon)$, and $\varepsilon_F$ denote the Boltzman's constant, the elementary charge, the electrical conductivity, and the Fermi energy, respectively. In the zero-energy limit, the thermopower of the free electron gas is related to its electronic specific heat by $S = C/ne$, where $C$ is the electronic specific heat and $n$ is the carrier concentration. Therefore, thermopower is highly sensitive to the change of Fermi surface and provides a measure of the entropy per itinerant carrier at low temperatures. The variations of thermopower can be used to characterize the nature of the QCP in heavy-fermion systems because the localized $f$-electrons do not participate in carrying the entropy, and their contribution to the thermopower is negligible around the QCP [17]. In the case of SDW quantum criticality, the *f*-electrons remain delocalized from below to above the QCP; thus, the thermopower is broadly symmetric around the QCP. However, in a Kondo-breakdown scenario, the localized-to-delocalized transition of *f*-electrons results in a pronounced asymmetry of the thermopower around the QCP. Here, we report a comparative study of the pressure-dependent thermopower of pure and Sn-doped CeRhIn$_5$. The striking difference in the thermopower behavior in the proximity to their respective QCP indicates the realization of different types of quantum criticality in pure and Sn-doped CeRhIn$_5$.

Single crystals of pure and Sn-doped CeRhIn$_5$ were synthesized using the In-flux method [18]. The quality of the crystals and the Sn concentration were confirmed via x-ray diffraction and energy-dispersive x-ray spectroscopy analysis, respectively. A hybrid piston-clamp-type cell was used to produce the quasi-hydrostatic pressure environment using Daphne oil as the pressure-transmitting medium. The pressure was determined from the pressure dependence of the superconducting transition temperature of Pb. Thermopower measurements were performed using a steady-state technique [19]. The heat current was injected parallel to the ab planes of the tetragonal structure, and the magnetic field was applied parallel to the heat flow. The generated temperature gradient $\Delta T$ was measured using a pair of chromel-Au$_{99.93\%}$Fe$_{0.07\%}$ thermocouples. The thermoelectric voltage $\Delta V$ was measured using a pair of voltage contacts aligned with the thermocouples. The thermopower was defined as $S = -\Delta V/\Delta T$. The electrical resistivity was measured by the same voltage contacts with additional current leads in a standard four-probe method. A HelioxVL system and a closed-cycle refrigerator were used in the temperature ranges of 0.25 to 10 K and 3 to 300 K, respectively.

Figure 1(a) shows the temperature dependence of the thermopower divided by temperature, $S/T$, for pure CeRhIn$_5$ at representative pressures, which was measured under a field of 8.8 T to reveal the thermopower behavior at low temperatures [11]. For $P < P_c$, e.g., at 1.26 GPa (blue symbols), the AFM transition ($T_N$) is characterized by an upturn in $S/T$, as denoted by the arrow, which tracks the transition temperature determined from the resistivity and heat-capacity measurements [11,12]. At temperatures above $T_N$, $S/T$ decreases abruptly upon cooling. As the pressure increases, the signature of $T_N$ weakens and eventually disappears at 1.75 GPa, whereas the abrupt decrease in $S/T$ remains up to 2.20 GPa. As the pressure increases further, e.g. at 2.30 GPa (brown symbols), $S/T$ increases monotonically with decreasing temperature down to the lowest temperature. The asymmetric behavior of $S/T$ around $P_c$ can be attributed to the localized-to-delocalized transition of *f*-electrons [13,14], as expected from the theoretical calculation for the Kondo-breakdown quantum criticality [17]. Another notable behavior is that the sign of $S/T$ changes from positive to negative at pressures below $P_c$. As shown in the contour plot of $S/T$ in Fig. 1(b), the temperature at which the sign changes, $T_{S/T=0}$, shifts to lower temperature with increasing pressure and finally terminates at $P_c$, which can be another signature of the Fermi-surface change at $P_c$. Further measurements at lower temperatures are needed to determine the sign of $S/T$, especially around $P_c$. The coincidence of the Fermi-surface change and magnetic criticality at $P_c$, where $T_c$ reaches its maximum, indicates that both the critical charge and magnetic quantum fluctuations mediate Cooper pairing in pure CeRhIn$_5$.

The temperature dependence of in-plane resistivity for Sn-doped CeRhIn$_5$ is shown in Fig. 2(a) at various pressures up to 2.0 GPa. Replacing 4.4% of the In atoms with Sn leads to a decrease of $T_N$ from 3.8 to 2.1 K at ambient pressure. The formation of the long-range AFM order is reflected by the inflection of the resistivity, as shown in the inset of Fig. 2(a) on the linear temperature scale. With the application of pressure, $T_N$ is gradually suppressed, and the pressure-induced superconductivity is observed. The pressure dependences of $T_N$ and $T_c$ are summarized in the temperature–pressure phase diagram in Fig. 2(b), superimposed with a color contour plot of resistivity. $T_N$ is extrapolated to $T = 0$ K at the upper critical pressure $P_{c2}$ (~1.3 GPa) where $T_c$ reaches a maximum. In the proximity to $P_{c2}$, as shown in Fig. 2(e), the resistivity exhibits a linear-temperature dependence, a hallmark of the non-Fermi liquid that typically emerges around a QCP. On the other hand, the Landau-Fermi-liquid behavior with the $T^2$ dependence is observed in the low- and high-pressure regimes, as displayed in Figs. 2(d) and 2(f). The results of the power-law analysis of the low-temperature resistivity at 7 T, i.e., $\rho = \rho_0 + AT^n$, are illustrated in Fig. 2(d). It demonstrates that the temperature coefficient, $A$, which is related to the effective mass of quasiparticles, peaks sharply at $P_{c2}$, further supporting the underlying QCP near $P_{c2}$. In addition to the QCP at $P_{c2}$, another critical pressure emerges at $P_{c1}$ (~1.0 GPa) where the residual resistivity $\rho_0$ reaches a maximum. This can be clearly observed from the contour plot of resistivity in Fig. 2 (b), which shows a funnel-shaped topology centered around $P_{c1}$. These anomalies provide evidence for the existence of two critical pressures, $P_{c1}$ ~ 1.0 GPa and $P_{c2}$ ~1.3 GPa, and the higher critical pressure $P_{c2}$ corresponds to an AFM QCP.

Figures 3(a) and 3(b) show the low-temperature thermopower $S/T$ of Sn-doped CeRhIn$_5$ for pressures around $P_{c1}$ and $P_{c2}$, respectively, where a magnetic field of 7 T was applied to suppress the superconductivity. Similar to the resistivity analysis, the two critical pressure points divide the system into three regions. In the pressure region below $P_{c1}$, the temperature dependence of $S/T$ is qualitatively comparable to that of pure CeRhIn$_5$ below $P_c$ and exhibits asymmetry around $P_{c1}$. Thus, it can be conjectured that the Fermi-surface reconstruction associated with the delocalization of $f$-electrons occurs at $P_{c1}$. The critical charge fluctuations originating from the Fermi-surface reconstruction results in the anomalous enhancement of residual resistivity around $P_{c1}$, as shown in Figs. 2(b) and 2(c). In contrast to the asymmetric behavior of S/T around $P_{c1}$, S/T increases symmetrically about $P_{c2}$ as it is approached from the AFM and paramagnetic states, as theoretically expected for the SDW quantum criticality [17]. In this case, the *f*-electrons remains itinerant and participates in carrying the entropy across the AFM QCP.

Following the theoretical model that considers the low-energy quasi-2D spin fluctuations associated with the AFM QCP [20], thermopower can be expressed as follows:

$$\frac{S}{T} \propto \frac{1}{e}\left(\frac{g_0^2 \mathcal{N}'(0)}{\varepsilon_F \omega_S \mathcal{N}(0)}\right) \ln\left(\frac{\omega_S}{\delta}\right),$$

where $\mathcal{N}(0)$ is the density of states at the Fermi energy $\varepsilon_F$, $g_0$ is the coupling between the electrons and spin fluctuations, $\omega_S$ is the typical energy of the spin fluctuations, and $\delta$ is the mass of the spin fluctuations. Since $\delta$ measures the deviation from the critical point, it can be written as $\delta = \Gamma(p - p_c) + T$, where $\Gamma$ is an appropriate energy parameter, and $p$ is an experimental parameter such as doping, pressure, or magnetic field, which can be tuned to the critical value $p_c$. In the regime where $T > \Gamma(p - p_c)$, $S/T \propto \ln(1/T)$, giving rise to the anomalous logarithmic temperature dependence of thermopower in the vicinity of the QCP. In our case, the S/T near $P_{c2}$ can be well described by this theoretical model and it increases logarithmically with decreasing temperature, as shown in Fig. 3(c). Being analogous to the non-Fermi liquid in resistivity, the logarithmic temperature dependence of S/T is another typical quantum-critical behavior that has been used as a distinctive signature of quantum criticality in many quantum-critical systems, including heavy fermions [21,22], cuprates [23,24], iron pnictide [25,26], cobalt oxides [27], and spin-liquid candidates [28]. With a further increase in pressure, the low-temperature S/T approaches a constant value, as exemplified for $P = 2.02$ GPa in Fig. 3(c), signifying the formation of the Fermi-liquid state. To clearly illustrate the overall profile of S/T, a color-contour plot of S/T is plotted in the pressure–temperature plane in Fig. 3(d). One immediately notices that S/T is highly enhanced around $P_{c2}$ and exhibits a fan-shaped dispersion. It is also shown in the pressure dependence of S/T at 0.6 K in Fig. 4(a), where S/T reaches a maximum near $P_{c2}$, and the rate of the change of S/T in the AFM phase is more pronounced than that in the paramagnetic phase because of the entropy reduction in the AFM ordered phase. S/T reflects the effective mass based on the relationship between S/T and C, S = C/ne, thus the maximum S/T is consistent with the enhancement of the temperature coefficient *A* near $P_{c2}$. Furthermore, the coefficient γ of the logarithmic temperature dependence of S/T obtained from the fits to the low-temperature regions using the relationship S/T = γln($T_0$/T) is a possible measure of the strength of the quantum-critical spin fluctuations [20]. As shown in Figs. 4(b) and 4(c), the critical fluctuations are the strongest near $P_{c2}$ where $T_c$ is a maximum. The correlation among the pressure dependence of S/T, *A*, γ, and $T_c$ suggests that the critical spin fluctuations mediate superconductivity in Sn-doped CeRhIn$_5$. We not that this correlation was in fact observed in iron pnictides [26], cuprates [23], and spin-liquid candidates [28].

In summary, the nature of quantum criticality in pure and Sn-doped CeRhIn$_5$ has been investigated by the systematic thermopower measurements under pressure. The $S/T$ is asymmetric around the AFM QCP in pure CeRhIn$_5$ but symmetric in the Sn-doped case, indicating that the nature of the magnetic criticality changes from the Kondo-breakdown type to the SDW type by 4.4% Sn substitution for In. Such a tuning effect on quantum criticality was also observed in the prototypical Kondo-breakdown system YbRh$_2$Si$_2$ by replacing Rh with a small amount of Co [29]. The change in the nature of quantum criticality can be elucidated using the global phase diagram, which describes the evolution of the ground state of heavy-fermion systems as a function of Kondo coupling and magnetic frustration [6,30]. As a function of the nonthermal tuning parameters (pressure or magnetic field in CeRhIn$_5$ and YbRh$_2$Si$_2$, respectively), the pure compounds follow the trajectory of a direct transition from an AFM state with a small Fermi surface to a paramagnetic state with a large Fermi surface, whereas the doped compounds follow a different trajectory that goes through an intermediate AFM state with a large Fermi surface before transitioning to a paramagnetic state with a large Fermi surface. Corresponding to the change in the quantum criticality, the quantum-critical fluctuations mediating the formation of Cooper pairs changes in pure and Sn-doped CeRhIn$_5$. These observations further support the shift of the charge-delocalization crossover scale revealed in previous Hall measurements [14] and suggest that thermopower can be an effective tool in elucidating quantum criticality and its connection to superconductivity in strongly correlated systems.

**Acknowledgments**

We acknowledge benefits from discussion with Prof K. Kim. This study was supported by the National Research Foundation (NRF) of Korea through a grant funded by the Korean Ministry of Science and ICT (Grant Nos. 2021R1A2C2010925, RS-2023-00220471, 2021R1I1A1A01047499, and 2022H1D3A3A01077468). Z.Y.C. also acknowledges the Special Construction Project Fund for Shandong Province Taishan Scholars.

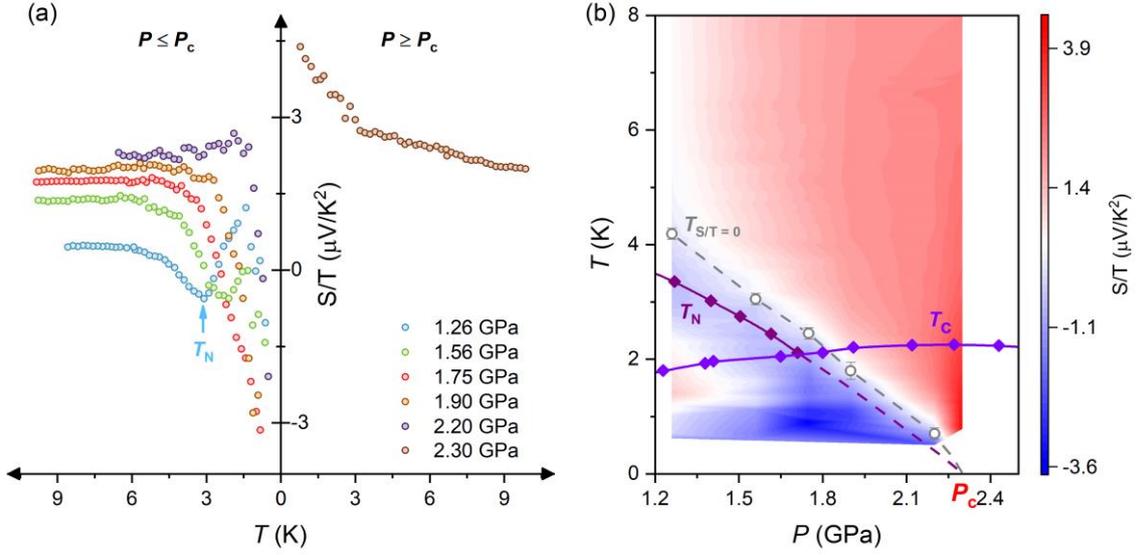

FIG. 1. (a) Temperature dependence of the thermopower divided by temperature $S/T(T)$ for CeRhIn$_5$ measured at 8.8 T and above the superconducting transition temperature $T_c$ for pressures less than $P_c$ (left panel) and greater than $P_c$ (right panel). The arrow indicates the AFM transition at $T_N$. (b) Contour plot of $S/T$ constructed from the data shown in Fig. 1(b). $T_c$ and $T_N$ are plotted in purple and violet symbols, respectively. The gray circles represent the temperature where $S/T = 0$. The dashed lines are guides to the eyes.

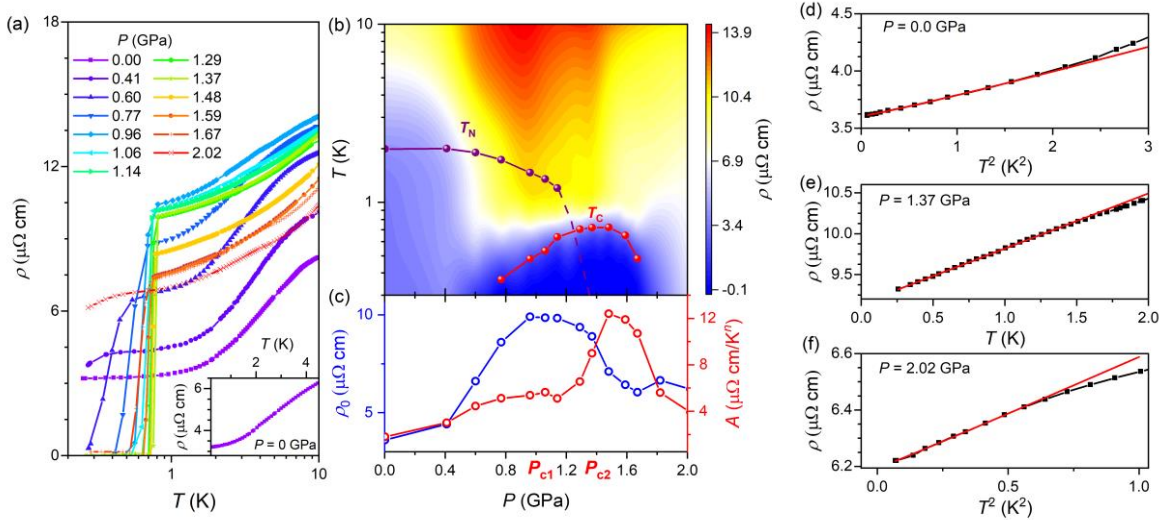

FIG. 2. (a) Temperature dependence of the in-plane resistivity $\rho(T)$ for 4.4% Sn-doped CeRhIn$_5$ at various pressures under a field of 0 T. The inset shows $\rho(T)$ at ambient pressure. (b) Contour plot of $\rho(T)$ constructed from the data shown in Fig. 2(a). $T_c$ and $T_N$ are denoted by purple and violet symbols,

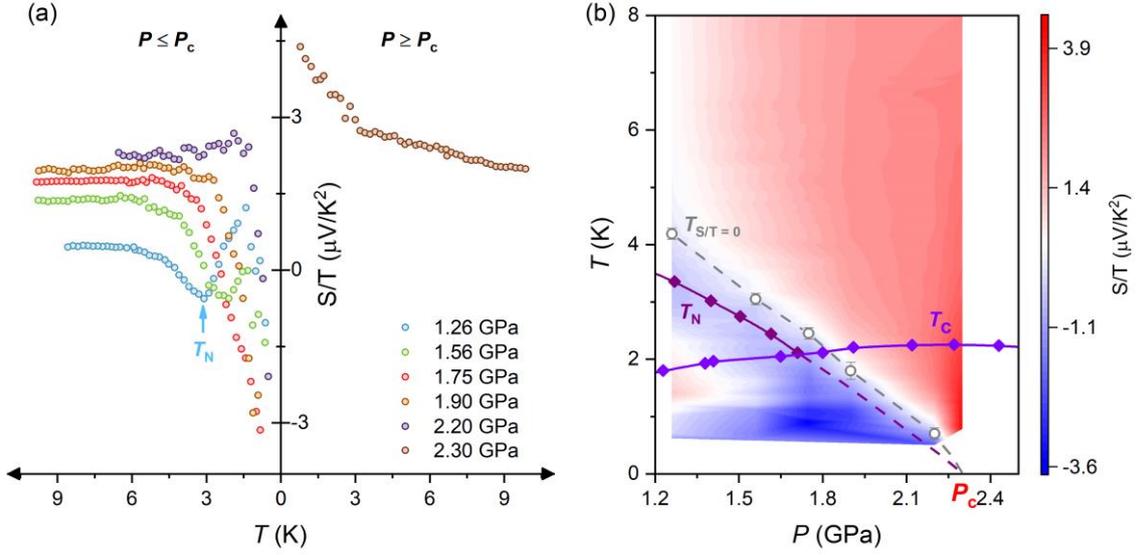

FIG. 1. (a) Temperature dependence of the thermopower divided by temperature $S/T(T)$ for CeRhIn$_5$ measured at 8.8 T and above the superconducting transition temperature $T_c$ for pressures less than $P_c$ (left panel) and greater than $P_c$ (right panel). The arrow indicates the AFM transition at $T_N$. (b) Contour plot of $S/T$ constructed from the data shown in Fig. 1(b). $T_c$ and $T_N$ are plotted in purple and violet symbols, respectively. The gray circles represent the temperature where $S/T = 0$. The dashed lines are guides to the eyes.

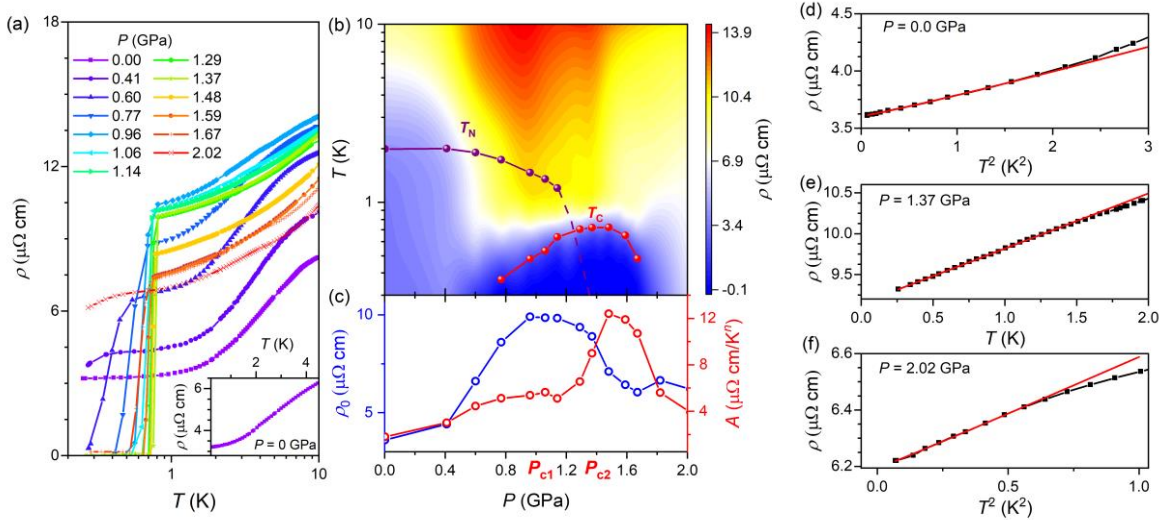

FIG. 2. (a) Temperature dependence of the in-plane resistivity $\rho(T)$ for 4.4% Sn-doped CeRhIn$_5$ at various pressures under a field of 0 T. The inset shows $\rho(T)$ at ambient pressure. (b) Contour plot of $\rho(T)$ constructed from the data shown in Fig. 2(a). $T_c$ and $T_N$ are denoted by purple and violet symbols,

respectively. (c) Pressure dependence of the temperature coefficient $A$ (right axis, red circles) and residual resistivity $\rho_0$ (left axis, blue circles) obtained from the power-law fits to the low-temperature resistivity at 7 T, i.e., $\rho_{ab} = \rho_0 + AT^n$. $P_{c1}$ (~1.0 GPa) indicates the critical pressure at which the residual resistivity $\rho_0$ is enhanced significantly. $P_{c2}$ (~1.3 GPa) represents the AFM QCP at which $T_N$ extrapolates to 0 K and $T_c$ reaches a maximum. (d)–(f) Low-temperature resistivity at 7 T for pressures of 0, 1.37, and 2.02 GPa, respectively. The resistivity shows a non-Fermi-liquid behavior near $P_{c2}$ but a Fermi-liquid behavior at low- and high-pressure regimes. The solid red lines are least-squares fits to the low-temperature resistivity.

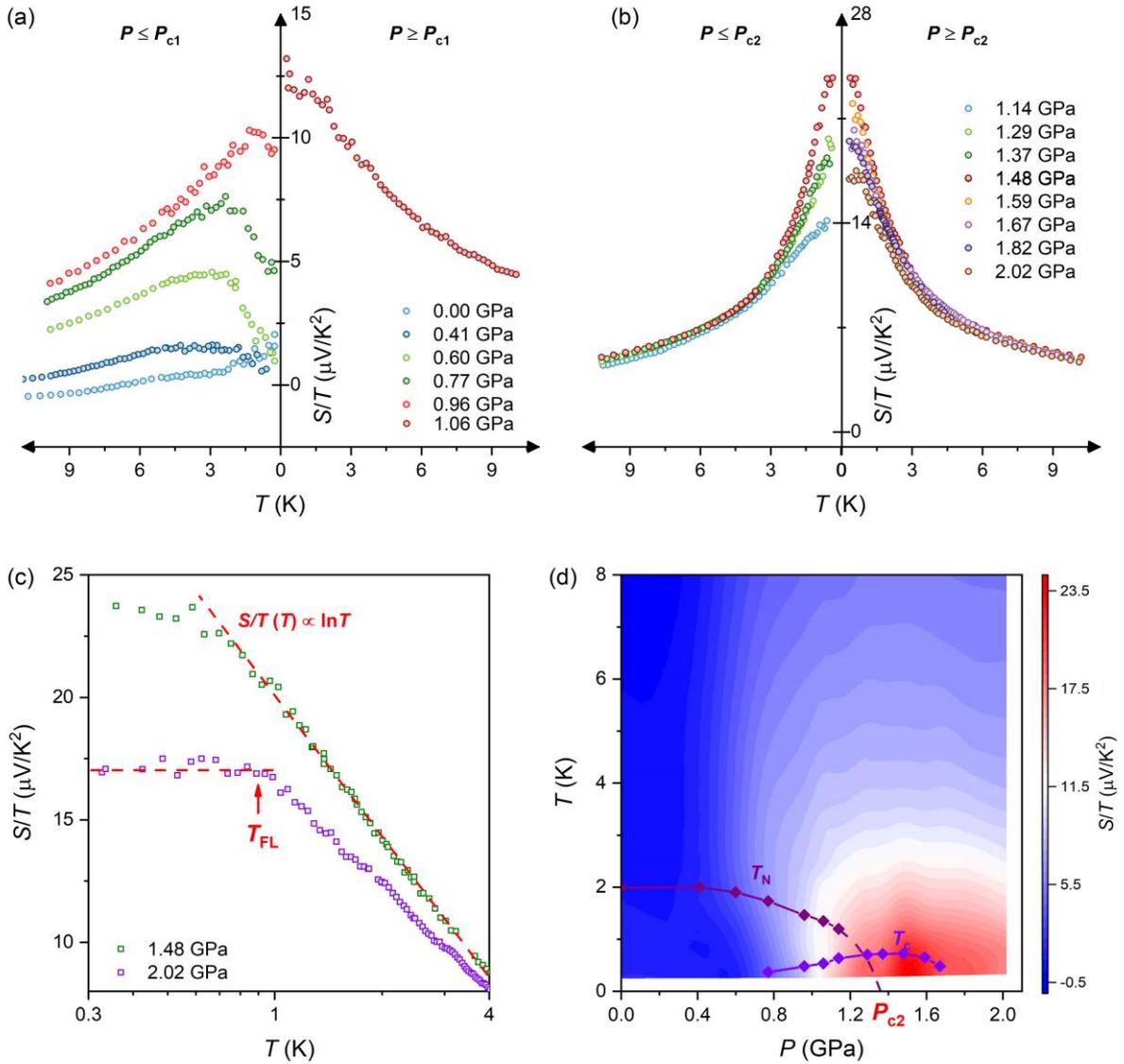

FIG. 3. Temperature dependence of the thermopower divided by temperature $S/T(T)$ for 4.4% Sn-doped CeRhIn$_5$ measured at 7.0 T for pressures around (a) $P_{c1}$ and (b) $P_{c2}$. (c) $S/T$ in the logarithmic

temperature scale at 1.48 and 2.02 GPa. Dashed red lines are guides to the eyes. (d) Contour plot of $S/T$ constructed from the data shown in Figs. 3(a) and 3(b). $T_c$ and $T_N$ are plotted in purple and violet symbols, respectively.

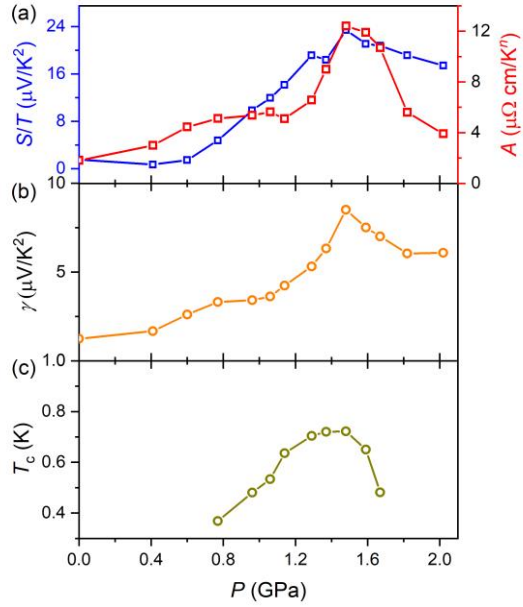

FIG. 4. Pressure dependence of the (a) $S/T$ at 0.6 K (left axis, blue squares) and temperature coefficient $A$ (right axis, red squares); (b) coefficient $\gamma$ and (c) superconducting transition temperature $T_c$ for 4.4% Sn-doped CeRhIn$_5$. $S/T$ at 0.6 K was obtained from the $S/T$ measured at 7 T, as shown in Fig. 3. The temperature coefficient $A$ was evaluated from the power-law fits to the low-temperature resistivity, $\rho_{ab} = \rho_0 + AT^n$, as shown in Fig. 2(c). The coefficient $\gamma$ was extracted from the fits of $S/T$ by $S/T = \gamma \ln(T_0/T)$. The superconducting transition temperature $T_c$ was determined by the zero-resistivity temperature, as shown in Fig. 2(a).